# Inverse design of broadband, strongly-coupled plexcitonic nonlinear metasurfaces

**Yael Blechman[1], Shai Tsesses[1], Guy Bartal[1] and Euclides Almeida[2, 3]**

[1]Andrew and Erna Viterbi Department of Electrical Engineering, Technion – Israel Institute of Technology, Haifa 32000, Israel
[2] Department of Physics, Queens College, City University of New York, Flushing, New York 11367, USA
[3] The Graduate Center of the City University of New York, New York, New York 10016, USA

Author e-mail address: Euclides.Almeida@qc.cuny.edu, guy@ee.technion.ac.il

## Abstract

Hybrid photonic structures of plasmonic metasurfaces coupled to atomically thin semiconductors have emerged as a versatile platform for strong light-matter interaction, supporting both strong coupling and parametric nonlinearities. However, designing optimized nonlinear hybrid metasurfaces is a complex task, as the multiple parameters' contribution to the nonlinear response is elusive. Here we present a simple yet powerful strategy for maximizing the nonlinear response of the hybrid structures based on evolutionary inverse design of the metasurface's near-field enhancement around the excitonic frequency. We show that the strong coupling greatly enhances the nonlinear signal, and that its magnitude is mainly determined by the Rabi splitting, making it robust to geometrical variations of the metasurface. Furthermore, the large Rabi splitting attained by these hybrid structures enables broadband operation over the frequencies of the hybridized modes. Our results constitute a significant step towards achieving flexible nonlinear control, which can benefit applications in nonlinear frequency conversion, all-optical switching, and phase-controlled nonlinear metasurfaces.

## Introduction

The field of artificial materials targets a constant search for the creation of structures with enhanced optical, electronic and other properties, enabling fine-grained control of light-matter interactions. In particular, hybrid structures consisting of plasmonic metasurfaces (MS) and atomically thin layers of transition metal dichalcogenides (TMDCs) have recently emerged as interesting and complex systems, where several nonlinear processes of different kinds exist simultaneously – a parametric nonlinearity and a strong cavity-emitter coupling.

Plasmonic MS are flat, subwavelength metallic nanostructures supporting surface plasmon polaritons (SPP), resulting from the coupling between transverse photons and longitudinal charge density propagating waves. Nanocavity arrays are particularly interesting kind of plasmonic MS, where different types of resonant behavior can co-exist, including localized and propagating SPPs, as well as cavity modes. These resonances create local field enhancement and confinement, while also enlarging the possibilities for fine-grained manipulation of light - in terms of directionality, phase, polarization etc., and design of novel functionality - lenses, holograms, modulators, all-optical switches, super-resolution and others[1–9] .

TMDC monolayers, on the other hand, are 2D direct bandgap semiconductors with unique mechanical, electronic, and optical properties[10], such as indirect to direct bandgap transition, enhanced mobility and conductivity, and mechanical flexibility, to name a few[11]. Their interaction with light is enhanced via the creation of Coulomb-bound electron-hole pairs, called excitons. The excitonic transitions in monolayer

TMDCs are in the visible/near-IR frequency range, typified by large binding energy and oscillator strength, and are stable at room temperature. These properties make them especially appealing for integration into optoelectronic devices.

When TMDC is coupled to a plasmonic MS, the latter acts as an open cavity, resulting in cavity-emitter coupling. The optical response of the TMDC can then be enhanced via the Purcell effect, increasing the interaction rate through a higher density of states. On the other hand, the cavity-emitter interaction can also form a strong coupling (SC) regime, where the two resonant modes exchange energy faster than the losses, resulting in the creation of new eigenmodes. These new eigenmodes are mixed light-matter states with modified coherence, propagation, and absorption properties, thereby creating even more tunability than either MS, excitons (in TMDC or molecular material[12–15]), or their weakly coupled combination. Realization of room-temperature SC in such hybrid systems is possible due to the substantial field enhancement and small mode volume that plasmonic cavities sustain despite the high associated losses, opening a thriving area of research with many possible applications[12,13,16–18].

On top of their unique attributes, both TMDCs and plasmonic MS can exhibit strong optical nonlinearity resulting from different mechanisms: the naturally high nonlinear susceptibility of noble metals in plasmonic MS, even further increased due to the local field enhancement[19,20], and the resonance of excitonic transitions in TMDCs [21–25], conveyed by enhanced oscillator strength when the exciton states serve as intermediate states for the nonlinear scattering process[21].

While SC in plasmonic-TMDC structures has been extensively studied in the linear regime, parametric nonlinear interactions in these systems are only starting to be explored, for second and third-harmonic generation in different setups, showing the possibility of significant NL enhancement[26–28]. However, combining several sub-diffraction-limit components usually results in complex behavior, especially when considering nonlinear processes. Therefore, obtaining nonlinear optimized structures is a computationally consuming task, usually requiring some inverse design process.

In this work, we develop a simple yet powerful strategy for broadband control over the nonlinear response in hybrid plexcitonic structures. We show that the response could be predicted by a single value of the Rabi splitting, and it is robust to variations in MS properties such as its geometry or the far-field linear response. We investigate the contribution of the coupling itself to the enhancement of Four-Wave Mixing (FWM) process, stressing the importance of the structures' near field response. Our method is based on inverse design of the metasurface's near-field enhancement, which consistently achieves significant strong cavity-emitter coupling.

## Results

We consider a metasurface consisting of a periodic array of nanocavities in a gold film with varying thickness, covered by a 10nm SiO$_2$ spacer. The excitonic layer is a 3nm-thick layer[1] of tungsten disulfide (WS$_2$ with the excitonic transition at $\lambda_{exc} = 612\ nm$) which is placed atop the spacer to form the hybrid structure (Figure 3a). Although exciton-photon interaction is essentially a quantum mechanical phenomenon, it can be described classically by a set of coupled harmonic oscillators[29,30] (Figure 1a). The new eigenmodes, the lower and upper polaritons (LP/UP), exhibit the typical anticrossing behavior with respect to some detuning parameter (Figure 1b). At the anticrossing point (when the resonant frequencies of the MS and the exciton are equal), the spectral distance between the eigenmodes equals the strength of the coupling (in units of energy). Therefore, the natural approach for designing such systems is to tailor the design of the plasmonic MS such that the plasmonic resonance matches the excitonic one. For a periodic array of nanoslits, the momentum matching condition and the plasmonic resonance location can be easily calculated using the SPP dispersion formula[31]. Figure 1c presents an example of a slit array configuration obtained using this "spectroscopic" approach, with Rabi splitting of 70 meV.

---

[1] Our reasons for choosing 3nm thickness were simulation mesh size and fabrication constraints (in practice flake transfer is possible for few monolayers and not a single monolayer). In our calculations we use the dielectric function of monolayer WS$_2$, which has a sharp resonant peak at the excitonic transition. Consequently, even though 3nm-thin film consists of 5-6 atomic layers, all our results below are approximately valid for 5 monolayers on top of each other. On the other hand, if a 3nm-thick bulk of WS$_2$ is transferred, we expect some deviations in practice.

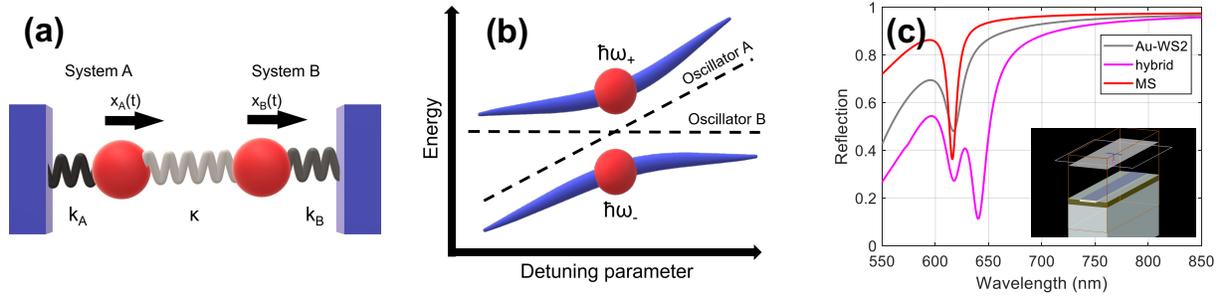

Figure 1 (a,b) The schematic representation of the strong coupling phenomenon, which can be described classically by a coupled oscillator model. (c) Linear reflection spectra of the uncoupled components and of the coupled hybrid system in the slit configuration obtained via the spectroscopic approach. Red shift in the resonances of the hybrid structure is caused by the change in the dielectric environment of the plasmon.

More complex structures, however, cannot be formulated analytically, hence, we have previously introduced a method for finding strongly coupled structures via a near-field inverse design approach[32]. In accordance with this method, we run an evolutionary optimization based on a genetic algorithm, where the optimization function is defined as the mean intensity $E_m$ of the near-field optical response in a small spectral window around $\lambda_{exc}$. The electric field is calculated using the finite-difference time-domain (FDTD) method along the incident polarization direction, as the excitons in the TMDC are excited mainly by in-plane components. In order to prioritize the creation of sharp, field enhanced features ("hot-spots"), we require that the minimal (normalized) field intensity used for the computation of $E_m$ be larger than a certain threshold. The method is described in details in Ref. [32]. We note that this procedure is analogous to minimizing mode volume in the expression $\Omega \sim \sqrt{N/V}$ for SC[29].

In order to demonstrate the generality of our approach for attaining broadband nonlinear control via strong coupling, we perform the optimization on several cavity geometries: arrays of doubly resonant rectangular cavities (dubbed as C1), of circular cavities (dubbed as C2) and singly resonant rectangular cavities (C3). In array C1, we modified the optimization function to provide a configuration that will both exhibit large SC and be resonant at the pump frequency, which should, in principle, contribute to the NL response. The corresponding hybrid structures exhibit Rabi splitting of ~170 meV for the optimized configurations and 70 meV for the slit configuration. The near-field design approach, shown in figure 2, results in structures with large localized field enhancements ("hot-spots"), and very small mode volumes, contributing to the large values of SC in these structures.

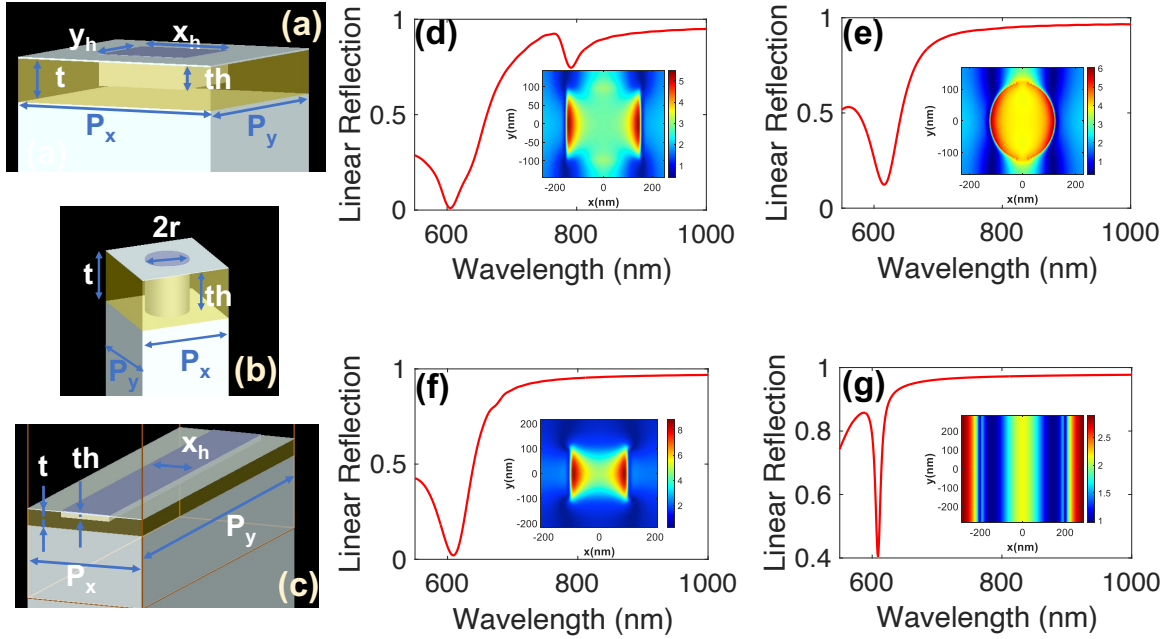

*Figure 2 The nanostructures designed to support SC. (a)-(c) The geometric optimization parameter spaces and (d)-(g) the corresponding linear reflection spectra of the near-field inverse designed nanocavity arrays (d-f) and the spectroscopically designed slits array (g). The near-field distribution of the $E_x$ component calculated at the spatial location of the WS$_2$ layer, in the MS structure, are shown as insets. The cavities/slits are milled in gold film of thickness t on top of a SiO$_2$ layer, the depth of the milling is th, the dimensions of the unit cell are Px and Py, with 10nm SiO$_2$ spacer on top of the gold. Geometric parameters of the configurations: (a,d) C1: doubly resonant rectangular cavity obtained by modifying the Em function, $x_h$ =300nm, $y_h$ =190nm, t=th=200nm, Px =500nm, Py =290nm. (b,e) C2: circular cavity, r=120nm, t=th=460nm, Px =460nm, Py =340nm (patterning through the spacer) (a,f) C3: singly resonant rectangular cavity, $x_h$ =$y_h$ =200nm, t=250nm, th=200nm, Px =420nm, Py =430nm. (c,g) nanoslit $x_h$=390 nm, t=100 nm, Px=Py=575nm, th=15nm.*

For each one of the four cavity geometries described above, we investigate the nonlinear response of both uncoupled MS and TMDC-coupled hybrid structures corresponding to these geometries. We calculate the far-field reflection spectra of a Four-Wave mixing (FWM) process - a general third-order nonlinear phenomenon where two collinear beams with frequencies $\omega_{Pump}$, $\omega_{Stokes}$ interact to produce a signal with frequency $\omega_{FWM} = 2\omega_{Pump} - \omega_{Stokes}$ (Figure 3a). The FWM signal is chosen to be at the excitonic resonance, 612 nm. This way, the generated FWM signal includes all the different contributions to nonlinearity, including plasmonic, excitonic, and modified nonlinear susceptibilities due to the coupling. The pump and Stokes wavelengths are 800nm and 1154nm, respectively. In Figure 3b we compare the nonlinear response of the bare MS, the TMDC layer on a bare gold film with the spacer (without patterning), and the hybrid structure (designated as "MS", "WS$_2$/Au" and "hybrid", respectively, as in Figure 3c) for the slit configuration. The nonlinear signal is obtained via nonlinear 3D FDTD simulations using a commercial Lumerical software package. The linear permittivity of WS$_2$ has been experimentally measured in Ref[33]. We assigned an excitonic contribution to the nonlinearity of WS$_2$ by a $\chi^{(3)}$ value of $2.4 \times 10^{-18} m^2/V^2$, reported for THG [34], which is the closest reported value in the literature for the signal at the excitonic A resonance. The response of the hybrid structure at $\omega_{FWM}$ was higher by a factor 5 compared to either the bare TMDC layer or the bare MS alone.

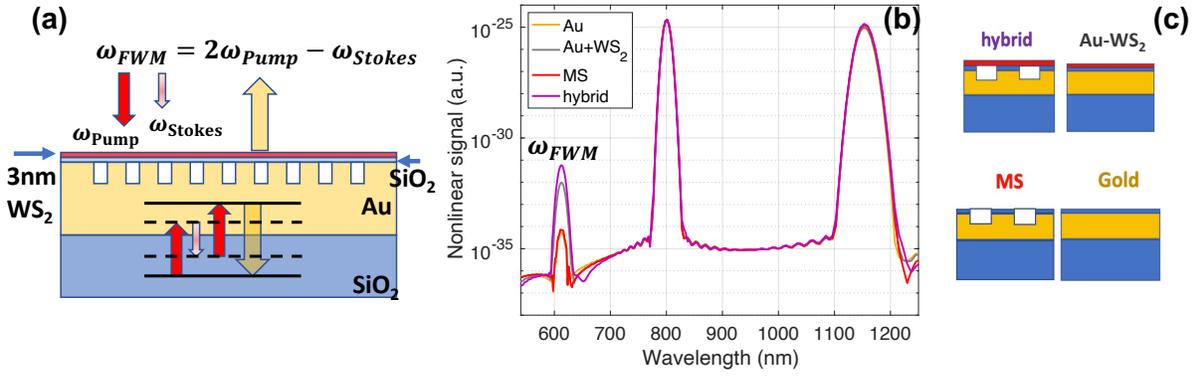

Figure 3 The hybrid structure and its different components, with their FWM response for the slits configuration. (a) The schematic representation of the FWM process. Two photons with frequency $\omega_{Pump}$ interact with a photon with frequency $\omega_{Stokes}$, producing a photon with frequency $\omega_{FWM}$. (b) Nonlinear spectra of uncoupled components and of the coupled hybrid system in the slits configuration: hybrid MS- $WS_2$, $WS_2$-Au (TMDC on top of a gold film without the patterning), MS and bare gold with spacer, clearly showing that the hybrid structure response is significantly larger than the response of each component separately.

We repeat the same comparison for the optimal configurations in Figure 4: the response of the optimized hybrid structure at $\omega_{FWM}$ was higher by over two orders of magnitude than the bare TMDC layer or the bare MS alone. To verify we are in the FWM nonlinear regime, we also show the dependence of the nonlinear signal on the pump field amplitude (keeping the Stokes field amplitude constant), obtaining 4th order dependence, as expected.

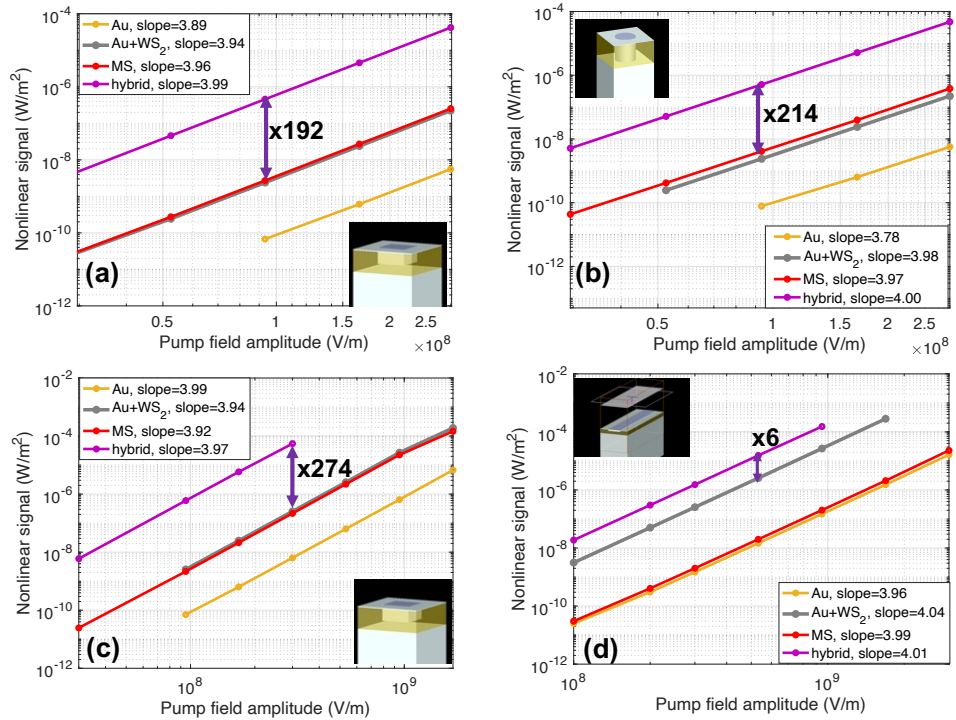

Figure 4 Nonlinear signal at the excitonic frequency. The computed value is the integral of the NL signal over the range of 30 nm around the center wavelength, normalized to the area of the unit cell. In each subplot, the response of hybrid, MS, Au-$WS_2$ and bare gold are shown as a function of pump field amplitude, exhibiting the fourth power dependence as expected in the FWM process. (a) C1, (b) C2, (c) C3, (d) slits.

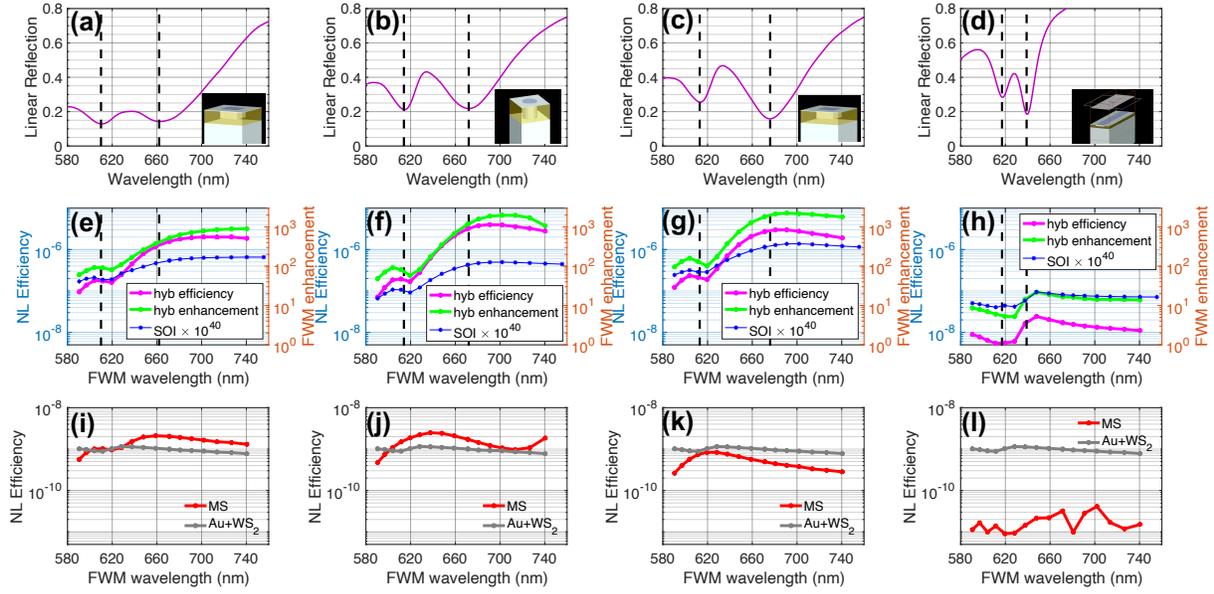

*Figure 5 Broadband linear and nonlinear response of the different configurations: (a,e,i) C1, (b,f,j) C2, (c,g,k) C3, (d,h,l) nanoslits. (a-d): linear reflection spectra with LP and UP denoted as black dashed lines. The spectral splitting is visible in all the configurations, see Figure 1 and compare with the MS spectra in Figure 2. (e-h): hybrid structure efficiency vs SOI and enhancement due to the coupling as a function of the FWM signal wavelength in the LP-UP range. The pump is kept fixed at 800nm. The efficiency value is computed as shown in the text. The SOI value is scaled to match the range of the enhancement values on the right axis. All simulations were done using the same source power of the pump and the Stokes beams. (i-l): the efficiency of the uncoupled components, MS and Au-WS$_2$ . In nanoslits array, the MS response is very weak, and as a result, the graph is noisy.*

The lower and upper polaritons span a wide range of frequencies in the strongly coupled system, opening up the prospect for broadband operation of nonlinear hybrid metasurfaces. In Figure 5, we examine the nonlinear response of all the structures, both in the hybrid (e-h, magenta curves, left axis) and the uncoupled form (i-l), when the FWM signal span the entire range of UP-LP frequencies by keeping the pump frequency fixed while changing the Stokes frequency $\omega_{Stokes}$ continuously. For each value of $\omega_{Stokes}$ we run the NL-FDTD simulation, obtain the nonlinear signal and compute the nonlinear efficiency as the ratio between the NL signal power in the far-field to the total input source power injected by the pump and the Stokes beams (and so, in particular, it is normalized to the unit cell area):

$$\text{efficiency} = \frac{\int_{FWM} R(f) df}{\int \text{SourcePower}(f) df}$$

$$\text{SourcePower}(f) = \frac{1}{2} \int_{\text{source injection plane}} \text{Re}\left(P(f)_{\text{Source}}(x,y)\right) \cdot dS$$

$$R(f) = \frac{1}{2} \int_{\text{far field monitor}} \text{Re}\left(P(f)_{\text{Signal}}(x,y)\right) \cdot dS$$

where $P(f)(x,y)$ denotes the Poynting vector. Evidently, the nonlinear efficiency of the hybrid structures is significantly higher than those of the bare MS or bare TMDC layer in the entire range between LP and UP (compare e-h and i-l in Figure 5). Interestingly, the linear resonance of C1 at the pump frequency did not result in significant increase of the NL efficiency compared to the other optimized configurations C2 and C3, showing the benefit of our approach of optimizing the SC response directly.

Another important observation is that the nonlinear efficiency of the coupled system and the linear response of the same system are correlated (compare a-d and e-h in Figure 5). Namely, the nonlinear efficiency of the optimized hybrid structures exhibits a local maximum at the spectral dips of the linear spectra, corresponding to the LP and UP frequencies. The nonlinear signal in the LP frequency is one order of magnitude larger than the one in the UP. This behavior can be explained by the differences in their nature (as we discuss in the next section) and absorption losses of the FWM signal by the gold film due to interband transitions, which are higher at the UP frequency. In contrast, the nonlinear efficiency plot of the spectroscopically designed configuration shows a peak only at the LP frequency. Note that the bare MS\TMDC layer exhibit a single peak at their respective resonances in all configurations (Figure 5 i-l).

One conclusion from the above results is that the addition of the nano-patterning is the cause of the significant NL enhancement. This enhancement includes the excitonic enhancement, the plasmonic enhancement (which we discussed in the introduction), and the enhancement due to the coupling. Since our optimization method produced strongly coupled configurations, it is interesting to isolate the pure contribution of the SC to the NL response. In order to do so, we take the values of the different efficiencies and define the coupling enhancement to be

$$Enhancement = \frac{efficiency(hybrid)}{efficiency(MS) + efficiency(Au + WS_2)}.$$

In Figure 5 (e-h, green curves, right axis), we plot these values for all our configurations as a function of the NL signal frequency. The maximal enhancement due to the coupling alone is above three orders of magnitude for the optimized configurations, while in the slits array configuration, there is only one order of magnitude contribution. This result correlates well with the SC values for these configurations: ~170 meV for the optimized ones and 70 meV for the slits. Furthermore, the nonlinear enhancement in the LP frequency for the slits configuration is larger than the one in the UP, while in the optimized configurations, this difference becomes even more prominent, reaching one order of magnitude (see also Supplementary Table T1, second column). This consistent difference is also present in the efficiency plots of the hybrid configurations, suggesting that the main contribution to the efficiency is the coupling – since the normalization by the response of the uncoupled components does not make a significant difference. This indicates the robustness of the SC influence on the nonlinearity to variations in the geometry of the structure and its linear response.

Note that both the emitter and the cavity are highly confined; therefore, their interaction occurs mainly in the near-field, which can be calculated to obtain the nonlinear response. Our approach for estimating the far-field nonlinear response in terms of the near field response is to consider the *linear* fields participating in the FWM process and compute the Spatial Overlap Integral (SOI) between these fields[19,35]:

$$NL \approx \iiint \chi^{(3)} E_{Pump}^2 E_{Stokes}^* E_{FWM} dV,$$

where $\chi^{(3)}$ is the nonlinear third-order susceptibility of the material. The fields $E_{Pump}, E_{Stokes}, E_{FWM}$ are the linear field amplitudes at the frequencies $\omega_{Pump}, \omega_{Stokes}, \omega_{FWM}$ respectively. All the variables depend on the spatial coordinates. Here, we extend this approach to strongly coupled NL hybrid structures. In Figure 5 (d-f, blue curve, right axis) we compute the value of the SOI for a sequence of different nonlinear processes, where $\omega_{FWM}$ varies in the range including LP and UP, and the $\omega_{Stokes}$ changed accordingly, so that the pump stays at 800nm. The results provide an excellent fit between the SOI and the FWM enhancement in the hybrid structures[2], showing that the NL response can be predicted from the **near-field** linear response -- while there is only a qualitative similarity with respect to far-field resonance locations.

Further influence of the strong coupling phenomenon on the nonlinear interaction can be seen in the near-field of the hybrid structures. In Figure 6 (see also Supplementary Figures S1-S2 for more details), we show the Poynting vector spatial distribution of the generated FWM signal at the LP and UP frequencies. The difference in far-field response is closely related to the localization of the hybridized modes, and in particular on the more delocalized nature of the UP mode. In addition, the nonlinear signal in the UP frequency is damped due to interband transitions in the metal[36].

## Summary

We introduced a simple yet powerful method for designing hybrid plasmonic-TMDC structures with high nonlinearities. This computationally inexpensive optimization method is based solely on optimizing the linear near-field of the metasurfaces at the excitonic frequency to produce configurations with strong coupling. We investigate the four-wave mixing response of the heterostructures and relate it to their coupling characteristics. The generated nonlinear signal includes all the coexisting contributions to the nonlinearity, namely plasmonic, excitonic, and modified nonlinear susceptibilities due to their coupling. We isolated each contribution in the optimized configurations and found a clear correlation between the value of the SC and the NL enhancement, which can reach up to three orders of magnitude.

The large Rabi splitting attained by these hybrid metasurfaces enables broadband nonlinear generation in the visible regime where the nonlinear enhancement is consistently much stronger at the lower polariton frequency. Using the near-field optimization approach, we found that the linear near-fields distribution of the hybridized modes determines the nonlinearity of the strongly coupled hybrid structures, while the far-field linear response alone is not sufficient for such a prediction.

We believe that our results constitute a step towards achieving versatile and robust nonlinear spectral control, which can in turn be very beneficial for a variety of applications.

---

[2] The specific ratios at the LP/UP frequencies are shown in Supplementary Table T1, first and third columns.

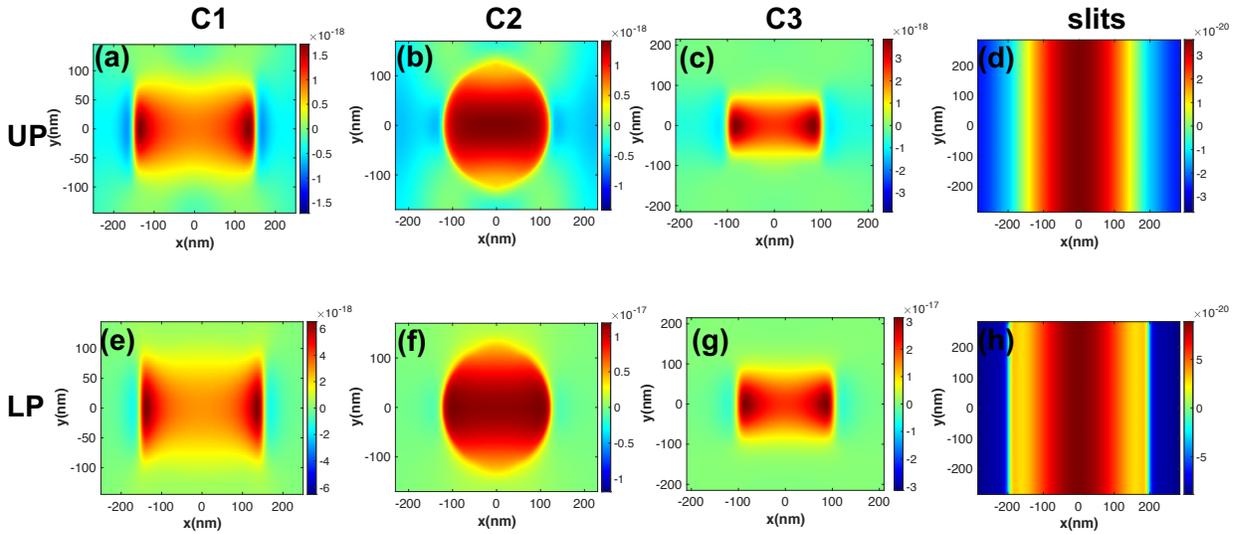

*Figure 6* The spatial distribution of the real part of the z component of the Poynting vector ($P_z$) at the WS$_2$-air interface when the FWM signal is either at UP (a-d) or LP (e-h) frequencies, in the unit cell of each one of the nanocavity array configurations. The LP mode is mostly localized – upward power flow above the hole and nearly zero above the metal. The increased delocalization and surface confinement of the UP mode is manifested by $P_z$ sign reversal between the hole and the metal. The maximal value in the LP mode is one order of magnitude larger than in the UP. In the slits configuration, the difference is less pronounced because the SC value is smaller.

## Acknowledgements

The authors would like to thank Prof. Ilya Goykhman for inspiring discussions.## References

# Supplementary information

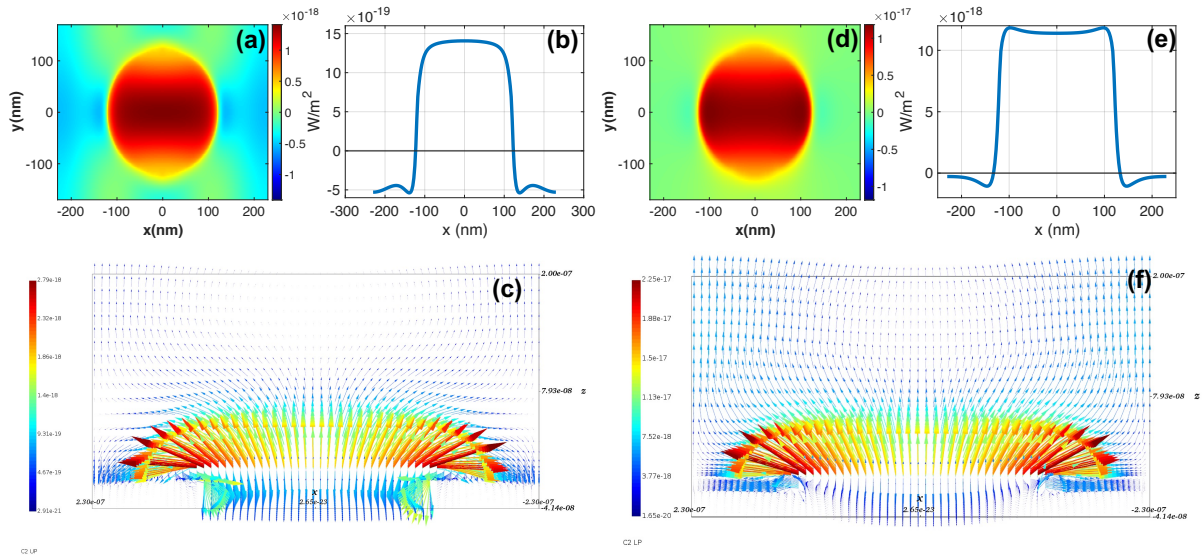

*Figure S1* FWM signal power flow in configuration C2. For the UP mode: (a) Spatial distribution of real part of Pz at the $WS_2$-air interface; (b) Horizontal slice of the data in (a) at horizontal cross-section (y=0); (c) Spatial distribution of the Poynting vector in the XZ plane at y=0. (d-f) same as (a-c) for the LP mode.

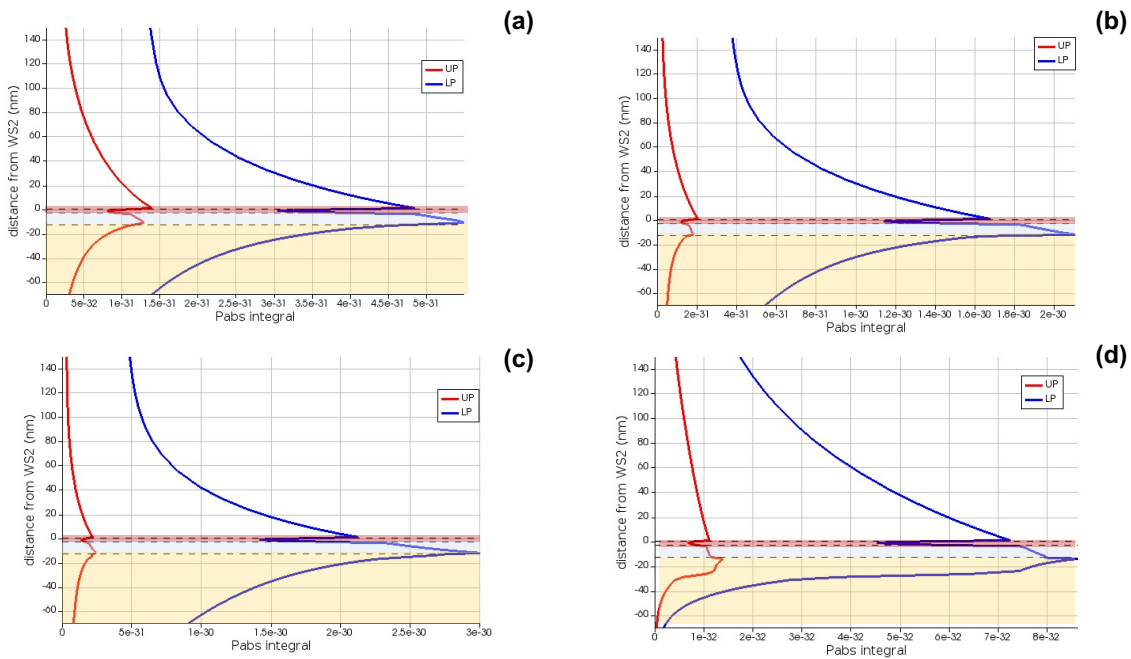

*Figure S2* FWM power as a function of distance from the interface in (a) C1, (b) C2, (c) C3, (d) nanoslits. The gold, $SiO_2$ and $WS_2$ layer boundaries are shown as dashed lines. The consistent difference between UP and LP can be seen in the different configurations.

|  | FWM enhancement LP/UP ratio | FWM efficiency LP/UP ratio | SOI LP/UP ratio |
| --- | --- | --- | --- |
| C1 | 4.2 | 7.0 | 2.6 |
| C2 | 15.2 | 17.0 | 4.3 |
| C3 | 15.6 | 19.7 | 4.8 |
| Slits | 2.7 | 3.1 | 1.3 |

*Table T1 Ratios between LP and UP enhancement/efficiency/SOI in the near field and the far-field, derived from the middle row of Figure 5.*